# Proving Termination of Normalization Functions for Conditional Expressions

Lawrence C Paulson

Computer Laboratory

University of Cambridge

3 June 1985

Boyer and Moore have discussed a recursive function that puts conditional expressions into normal form [1]. It is difficult to prove that this function terminates on all inputs. Three termination proofs are compared: (1) using a measure function, (2) in domain theory using LCF, (3) showing that its *recursion relation*, defined by the pattern of recursive calls, is well-founded. The last two proofs are essentially the same though conducted in markedly different logical frameworks. An obviously total variant of the normalize function is presented as the 'computational meaning' of those two proofs.

A related function makes nested recursive calls. The three termination proofs become more complex: termination and correctness must be proved simultaneously. The recursion relation approach seems flexible enough to handle subtle termination proofs where previously domain theory seemed essential.

# Contents

# 1 A normalization function

Boyer and Moore have published a machine-assisted proof of the correctness of a tautology checker for propositional logic [1]. Propositions are represented as conditional expressions (henceforth *expressions*). An expression is either an *atom* $At(a)$ for some symbol $a$, or else has the form $If(x, y, z)$ where $x$, $y$, $z$ are themselves expressions. An atom represents a propositional letter, while $If(x, y, z)$ is equal to $y$ if $x$ is true, and equal to $z$ otherwise.

The tautology checker includes a function for putting expressions into normal form. An expression is in *normal form* if it has no *tested Ifs*: subexpressions of the form $If(If(u, v, w), y, z)$. Replacing this by $If(u, If(v, y, z), If(w, y, z))$ preserves the value of the entire expression and removes one tested *If*. However new tested *Ifs* are created whenever $v$ or $w$ begin with *If*.

The following program, written in Standard ML [6], defines the data structure *exp* and the normalize function *norm*. If its argument is a tested *If* then *norm* replaces it as above and calls itself recursively. For any other argument *norm* makes recursive calls on the subexpressions. The ML code should fill in the details:

**type rec** $exp =$ **data** $At$ **of** $string \mid If$ **of** $exp \times exp \times exp$;

**fun** $norm(At(a)) = At(a) \mid$
$\qquad norm(If(At(a), y, z)) = If(At(a), norm(y), norm(z)) \mid$
$\qquad norm(If(If(u, v, w), y, z)) = norm(If(u, If(v, y, z), If(w, y, z)));$

It is far from obvious that *norm* terminates. In the *If-If* case it calls itself with a larger expression than it was given. One way of proving termination is to find a *well-founded relation* under which the argument 'goes down' in every recursive call [1, 5]. Classically, a relation $\prec$ is well-founded if and only if it has no infinite descending chains $\cdots \prec x_2 \prec x_1 \prec x_0$. The less-than relation $<$ on the set $\mathbf{N}$ of natural numbers is well-founded. Less-than is not well-founded on certain other sets: for the integers, $\cdots < -2 < -1 < 0$, and for the rationals, $\cdots < .01 < .1 < 1$.

A common way of defining a well-founded relation on a set $A$ uses a *measure function* $f : A \to \mathbf{N}$, defining $a' \prec a \iff f(a') < f(a)$. Then $\prec$ is the *inverse image* of $<$ under $f$. The *lexicographic combination* of two well-founded relations $\prec_A$ and $\prec_B$ defines a well-founded relation $\prec$ on pairs $\langle a, b \rangle$. Here $\langle a', b' \rangle \prec \langle a, b \rangle$ if and only if $a' \prec_A a$ or $a' = a$ and $b' \prec_B b$.

## 2   A proof using a measure function

In Boyer and Moore's logic all functions are total. Their theorem prover only accepts a recursive definition if it can show that the function terminates on all arguments. For this purpose it uses well-founded relations consisting of lexicographic combinations of inverse images. Boyer and Moore present a well-founded relation for *norm* involving two measures on expressions. Boyer has also sent me a simpler proof, credited to R. Shostak, using a single measure function:

**fun**   $m(At(a)) = 1 \mid$
$\quad m(If(x,y,z)) = m(x) + m(x) \times m(y) + m(x) \times m(z).$

To show that this measure goes down in each of *norm*'s recursive calls is a tedious exercise of expanding and collecting terms. It is important to check the easy *If-At* case, because a clever measure that goes down in the hard *If-If* case may not go down in the easy case. Note that $m(x)$ is positive for all $x$.

Let $U = m(u)$, $V = m(v)$, etc. The *If-At* case terminates because $Y < 1+Y+Z$ and $Z < 1+Y+Z$. For the *If-If* case the recursive call has measure

$$m(If(u, If(v,y,z), If(w,y,z)))$$
$$= U + U(V + VY + VZ) + U(W + WY + WZ)$$
$$= U + UV + UVY + UVZ + UW + UWY + UWZ$$

and the original argument has measure

$$m(If(If(u,v,w),y,z))$$
$$= U + UV + UW + (U + UV + UW)Y + (U + UV + UW)Z$$
$$= U + UV + UW + UY + UVY + UWY + UZ + UVZ + UWZ \ .$$

Cancelling common terms, this case terminates because $UY + UZ > 0$.

## 3   A proof in the Logic of Computable Functions

Jacek Leszczyłowski [4] has proved the termination of *norm* using the theorem prover Edinburgh LCF [3]. LCF's logic, a formalization of domain theory, allows reasoning about partial functions. Leszczyłowski's proof uses a lemma that the termination of *norm* in particular cases implies termination in other cases.

Each domain contains an 'undefined' element $\bot$, representing the result of a divergent computation. There is a *weak equality* predicate $\equiv$ such that $x \equiv y$ iff $x$ and $y$ are both undefined, or both defined and equal. Since quantifiers often range over defined values only, let $\forall_D x. P(x)$ abbreviate $\forall x. x \not\equiv \bot \Rightarrow P(x)$. The statement

'*norm* is total' is expressed as $\forall_D x. norm(x) \not\equiv \bot$. The domain of expressions is *flat* to avoid having infinite expressions [9]. The constructor functions *At* and *If* are total:

$$\forall_D a.\, At(a) \not\equiv \bot \qquad \forall_D xyz.\, If(x,y,z) \not\equiv \bot$$

Structural induction for expressions is

$$\frac{\forall_D a.\, P(At(a)) \qquad \forall_D xyz.\, P(x) \wedge P(y) \wedge P(z) \Rightarrow P(If(x,y,z))}{\forall_D x.\, P(x)}$$

This rule is often stated with the additional premise $P(\bot)$. Then the conclusion is $\forall x.\, P(x)$.

The function *norm* is expressed as three equations in LCF:

$$\begin{aligned}
\forall_D a.\, norm(At(a)) &\equiv At(a) \\
\forall_D ayz.\, norm(If(At(a), y, z)) &\equiv If(At(a), norm(y), norm(z)) \\
\forall_D uvwyz.\, norm(If(If(u,v,w), y, z)) &\equiv norm(If(u, If(v,y,z), If(w,y,z)))
\end{aligned}$$

The termination proof involves a lemma that if $norm(y)$ and $norm(z)$ terminate, then $norm(If(x,y,z))$ terminates.

**Lemma.** $\forall_D xyz.\, norm(y) \not\equiv \bot \wedge norm(z) \not\equiv \bot \Rightarrow norm(If(x,y,z)) \not\equiv \bot$

**Proof.** By structural induction on $x$. The *At* case reduces to showing

$$\forall_D ayz.\, norm(y) \not\equiv \bot \wedge norm(z) \not\equiv \bot \Rightarrow If(At(a), norm(y), norm(z)) \not\equiv \bot$$

which follows because *If* and *At* are total.

The *If* case reduces to showing

$$norm(y) \not\equiv \bot \wedge norm(z) \not\equiv \bot \Rightarrow norm(If(u, If(v,y,z), If(w,y,z))) \not\equiv \bot$$

assuming that $u, v, w, y, z$ are all defined and with induction hypotheses for $u$, $v$, and $w$:

$$\begin{aligned}
\forall_D yz.\, norm(y) \not\equiv \bot \wedge norm(z) \not\equiv \bot &\Rightarrow norm(If(u,y,z)) \not\equiv \bot \\
\forall_D yz.\, norm(y) \not\equiv \bot \wedge norm(z) \not\equiv \bot &\Rightarrow norm(If(v,y,z)) \not\equiv \bot \\
\forall_D yz.\, norm(y) \not\equiv \bot \wedge norm(z) \not\equiv \bot &\Rightarrow norm(If(w,y,z)) \not\equiv \bot
\end{aligned}$$

Assume $norm(y) \not\equiv \bot$ and $norm(z) \not\equiv \bot$. The induction hypotheses for $v$ and $w$ imply

$$norm(If(v,y,z)) \not\equiv \bot \quad \text{and} \quad norm(If(w,y,z)) \not\equiv \bot\,.$$

Instantiate the induction hypothesis for $u$ with $y \to If(v,y,z)$ and $z \to If(w,y,z)$, proving

$$norm(If(u, If(v,y,z), If(w,y,z))) \not\equiv \bot\,.$$

Q.E.D.

Termination of *norm* on all inputs follows by induction in $\forall_D x. norm(x) \not\equiv \bot$.

# 4 Proving the recursion relation is well-founded

The first termination proof defines a well-founded relation using a measure function, and shows that *norm*'s recursive calls obey that relation. A dual approach is to define a relation $\prec$ in terms of *norm*'s recursive calls, then show that $\prec$ is well-founded. Define $x \prec y$ to be *true* whenever evaluating $norm(x)$ requires a recursive call $norm(y)$, and to be *false* otherwise. (It should be *false* whenever possible, since additional relationships between elements could prevent $\prec$ from being well-founded.) I call $\prec$ the *recursion relation* of *norm*.

Case analysis of *norm* defines its recursion relation:

$$\begin{aligned} x \prec At(a) &\iff \textit{false} \\ x \prec \textit{If}(At(a), y, z) &\iff x = y \lor x = z \\ x \prec \textit{If}(\textit{If}(u, v, w), y, z) &\iff x = \textit{If}(u, \textit{If}(v, y, z), \textit{If}(w, y, z)) \end{aligned}$$

To show the termination of *norm* it suffices to show that the relation $\prec$ is well-founded. This proof will have a remarkable similarity to the LCF proof, which was conducted in domain theory. This section uses a simple mathematical framework with no partial elements. It uses *constructive* mathematics because this paper is an outgrowth of my study of well-founded relations [10] in Martin-Löf's *Constructive Type Theory* [8].

Showing that a relation is well-founded requires showing the soundness of its rule of *well-founded induction* for an arbitrary predicate $P$:

$$\frac{\forall x.\,(\forall x'.\,x' \prec x \Rightarrow P(x')) \Rightarrow P(x)}{\forall x.\,P(x)}$$

In constructive reasoning, showing that $\prec$ has no infinite descending chains is insufficient to verify the rule. The rule is verified directly, proving its conclusion from its premise. For the rest of this section assume the *induction step*:

$$\forall x.\,(\forall x'.\,x' \prec x \Rightarrow P(x')) \Rightarrow P(x) \tag{1}$$

Termination follows from proving $\forall x. P(x)$; a lemma is helpful.
**Lemma.** $\forall xyz.\,P(y) \land P(z) \Rightarrow P(\textit{If}(x, y, z))$
**Proof.** By structural induction on $x$. The *At* case is

$$\forall ayz.\,P(y) \land P(z) \Rightarrow P(\textit{If}(At(a), y, z))\;,$$

which follows from (1) and the definition of $\prec$. Recall that $\textit{If}(At(a), y, z)$ has only two predecessors, $y$ and $z$.

The *If* case is $\forall yz. P(y) \wedge P(z) \Rightarrow P(If(If(u,v,w),y,z))$ under the induction hypotheses
$$\forall yz. P(y) \wedge P(z) \Rightarrow P(If(u,y,z))$$
$$\forall yz. P(y) \wedge P(z) \Rightarrow P(If(v,y,z))$$
$$\forall yz. P(y) \wedge P(z) \Rightarrow P(If(w,y,z)) \ .$$

By (1) it is enough to show
$$P(y) \wedge P(z) \Rightarrow P(If(u, If(v,y,z), If(w,y,z))) \ .$$

Assume $P(y)$ and $P(z)$. The induction hypotheses for $v$ and $w$ imply $P(If(v,y,z))$ and $P(If(w,y,z))$. Instantiate the induction hypothesis for $u$ with $y \to If(v,y,z)$ and $z \to If(w,y,z)$, proving $P(If(u, If(v,y,z), If(w,y,z)))$. Q.E.D.

Now $\forall x.P(x)$ follows immediately by induction on $x$.

The previous proof can be translated into this one by replacing $norm(x) \not\equiv \bot$ by $P(x)$. Each unfolding of *norm* becomes an appeal to the induction step (1). Perhaps domains and partial objects are not essential even for difficult proofs of termination.

As a sample proof in his higher-order theory of constructions, Thierry Coquand has proved the termination of normalization [2] (pages 46–48). He defines a predicate $N(x)$ to mean '$x$ can be put into normal form,' and proves $\forall x. N(x)$. Translated from his formalism, the axioms are

$$N(At(a))$$
$$N(y) \wedge N(z) \Rightarrow N(If(At(a),y,z))$$
$$N(If(u, If(v,y,z), If(w,y,z))) \Rightarrow N(If(If(u,v,w),y,z))$$

The connection between $N(x)$ and $norm(x) \not\equiv \bot$ is obvious. Coquand builds a proof object resembling my Constructive Type theory one, using a similar Lemma.

## 5 An obviously total normalize function

Constructive Type Theory provides a formal interpretation of propositions as types. One consequence is that every proof by induction involves constructing a proof object by recursion. My Type Theory proof that $\prec$ is well-founded suggests another way of writing the normalize function:

**fun** $normif(At(a),y,z) = If(At(a),y,z)$ |
  $normif(If(u,v,w),y,z) = normif(u, normif(v,y,z), normif(w,y,z))$;

**fun** $norm_1(At(a)) = At(a)$ |
  $norm_1(If(x,y,z)) = normif(x, norm_1(y), norm_1(z))$;

The function *normif* is obviously total because it is structural recursive in its first argument, a sort of 'higher type' recursion. Although *normif* makes nested recursive calls in its second and third arguments, these have no effect on termination. (Ackermann's function is another example where termination is obvious despite nested recursive calls.) Note the similarity between *normif*'s recursive calls and the appeals to the induction hypotheses in the proof of the lemma.

Proving in LCF that $\forall_D x. \, norm(x) \equiv norm_1(x)$ constitutes yet another termination proof for *norm*. Our familiar lemma now takes the form

$$\forall_D xyz. \, norm(\mathit{If}(x,y,z)) \equiv normif(x, norm(y), norm(z)) \, ,$$

with essentially the same proof as before.

There is a pleasing concreteness about the first termination proof. But the measure function offers little intuition. The second and third proofs convey something of what *norm* is actually doing, for they give us the function *normif*.

## 6 A normalize function with nested recursion

If we modify the *If-If* case of *norm* to make nested recursive calls, proving termination becomes trickier still. Call the new function $norm_2$:

**fun** $norm_2(At(a)) = At(a) \mid$
  $norm_2(\mathit{If}(At(a), y, z)) = \mathit{If}(At(a), norm_2(y), norm_2(z)) \mid$
  $norm_2(\mathit{If}(\mathit{If}(u,v,w), y, z)) =$
    $norm_2(\mathit{If}(u, norm_2(\mathit{If}(v,y,z)), norm_2(\mathit{If}(w,y,z))));$

I sent this function as a challenge to Boyer and Moore. The version of the theorem prover described in their book [1] cannot handle this nested recursion. It could not admit $norm_2$ as a function unless, for some measure $m_2(x)$, it could prove

$$m_2(\mathit{If}(u, norm_2(\mathit{If}(v,y,z)), norm_2(\mathit{If}(w,y,z)))) \; < \; m_2(\mathit{If}(\mathit{If}(u,v,w), y, z)) \, .$$

Yet this very statement involves $norm_2$.

Moore informs me that the theorem prover has since been extended. A nested recursive function definition can be admitted by showing that it is equivalent to some already accepted definition. In this case replace $norm_2$ by *norm* in the recursion equations and show that the new equations hold. Thus they have at least one solution: *norm*. Then show that some measure decreases for each recursive call of *norm* in the new equations. Thus the solution is unique: by well-founded induction on the measure, $norm(x) = norm_2(x)$ for all $x$. Reasoning about *norm* is possible because it is already known to be a total function. Moore describes this principle of definition in his paper on the termination of Takeuchi's function [7].

Moore's proof of $norm_2$ has several stages:

- $norm(norm(x)) = norm(x)$ is proved by induction on the measure $m(x)$. The result is used in the $At$ case of the next stage.

- $\forall yz.\, norm(If(x, norm(y), norm(z))) = norm(If(x, y, z))$ is proved, like the Lemma, by structural induction on $x$. The theorem prover does not allow quantified induction schemes, but any instance of one can be specified.

- Therefore $norm$ is a solution to the equations for $norm_2$.

- A function to count the number of tested *Ifs* in an expression is defined. This differs from the function `IF.DEPTH` of the original proof [1], which counts the nesting of tested *Ifs*.

- $norm(x)$ is indeed normal: it contains no tested *Ifs*. Proved by induction on $m(x)$.

- The measure for proving uniqueness is the lexicographic combination of the number of tested *Ifs* and the size of an expression.

## 7 The LCF proof revisited

In domain theory the termination of $norm_2$ can be proved without any mention of $norm$. Termination and partial correctness must be proved *simultaneously*. Showing termination of the *If-If* case requires showing that the nested calls yield normal expressions.

Define the predicate $ISN(x)$ to hold whenever $x$ is in normal form. $ISN$ is a recursive predicate but the recursion is trivially well-founded:

$$\begin{aligned}
ISN(\bot) &\iff false \\
\forall_D a.\, ISN(At(a)) &\iff true \\
\forall_D ayz.\, ISN(If(At(a), y, z)) &\iff ISN(y) \wedge ISN(z) \\
\forall_D uvwyz.\, ISN(If(If(u, v, w), y, z)) &\iff false
\end{aligned}$$

The element $\bot$ is not in normal form under this definition; $\forall_D x.\, ISN(norm_2(x))$ states that $norm_2$ is a total function whose result is always normal. This is not a complete statement of correctness; it mentions no relationship between $x$ and $norm_2(x)$.

The proof resembles that of section 3, replacing each occurrence of $norm(x) \not\equiv \bot$ by $ISN(norm_2(x))$.

**Fact.** If the argument of $norm_2$ is normal then so is its result:

$$\forall x. ISN(x) \Rightarrow ISN(norm_2(x)) \qquad (2)$$

**Proof.** By structural induction on $x$. The $\bot$ and $At$ cases are easy. For $If$ consider two cases. Since $If(If(u,v,w),y,z)$ is not normal the result holds vacuously. The $If(At(a),y,z)$ case is

$$ISN(If(At(a),y,z)) \Rightarrow ISN(norm_2(If(At(a),y,z)))$$

which simplifies to

$$ISN(y) \wedge ISN(z) \Rightarrow ISN(norm_2(y)) \wedge ISN(norm_2(z))$$

which follows from the induction hypotheses.

Now we have the usual

**Lemma.** $\forall_D xyz.\ ISN(norm_2(y)) \wedge ISN(norm_2(z)) \Rightarrow ISN(norm_2(If(x,y,z)))$

**Proof.** By structural induction on $x$. The $At$ case reduces to the clearly true

$$\forall_D ayz.\ ISN(norm_2(y)) \wedge ISN(norm_2(z)) \Rightarrow ISN(If(At(a), norm_2(y), norm_2(z)))$$

The $If$ case reduces to showing, under induction hypotheses,

$$ISN(norm_2(y)) \wedge ISN(norm_2(z)) \Rightarrow ISN(norm_2(If(u, If(v,y,z), If(w,y,z)))) \ .$$

Assume $ISN(norm_2(y))$ and $ISN(norm_2(z))$. The induction hypotheses for $v$ and $w$ imply $ISN(norm_2(If(v,y,z)))$ and $ISN(norm_2(If(w,y,z)))$. Now comes a clear departure from the section 3 proof: inserting an extra call to $norm_2$. The Fact (2) gives

$$ISN(norm_2(norm_2(If(v,y,z)))) \quad \text{and} \quad ISN(norm_2(norm_2(If(w,y,z)))) \ .$$

Instantiate the induction hypothesis for $u$ with $y \to norm_2(If(v,y,z))$ and $z \to norm_2(If(w,y,z))$. Q. E. D.

Again the overall proof for $norm_2$ is an easy induction using the Lemma. Proving $\forall_D x.ISN(norm_2(x))$ rather than $\forall_D x.norm_2(x) \not\equiv \bot$ is a classic example of strengthening the goal in order to strengthen the induction hypotheses. The Lemma is essentially the inductive step. Its proof requires the Fact (2). Proving $\forall_D x.norm_2(x) \not\equiv \bot$ would require a Lemma of the form

$$\forall_D xyz.\ norm_2(y) \not\equiv \bot \wedge norm_2(z) \not\equiv \bot \Rightarrow norm_2(If(x,y,z)) \not\equiv \bot$$

and a Fact of the form $\forall x. x \not\equiv \bot \Rightarrow norm_2(x) \not\equiv \bot$. We are going in circles! It would suffice to prove a weaker version of the Fact:

$$\forall x. norm_2(x) \not\equiv \bot \Rightarrow norm_2(norm_2(x)) \not\equiv \bot .$$

An attempted proof of this resembles that of (2) except that the *If-If* case is no longer trivial.

## 8  The recursion relation proof revisited

The recursion relation proof of section 4 can similarly be adapted to $norm_2$. I continue to use the predicate *ISN* for reasoning about normal expressions, though in this section there is no element $\bot$. The recursion relation $\prec_2$ is defined like $\prec$, except that the *If-If* case has three recursive calls instead of one. The outer call involves the results of the inner calls, expressed as existentially quantified variables. The results are *assumed* to be in normal form:

$$
\begin{aligned}
x \prec_2 At(a) &\iff \textit{false} \\
x \prec_2 \textit{If}(At(a), y, z) &\iff x = y \lor x = z \\
x \prec_2 \textit{If}(\textit{If}(u, v, w), y, z) &\iff \left( \begin{array}{l} x = \textit{If}(v, y, z) \lor \\ x = \textit{If}(w, y, z) \lor \\ \exists v' w'. \textit{ISN}(v') \land \textit{ISN}(w') \land x = \textit{If}(u, v', w') \end{array} \right)
\end{aligned}
$$

It will be necessary to show inductively that the equations for $norm_2$ produce normal expressions. First let us show that $\prec_2$ is well-founded. Assume the induction step for an arbitrary $P$:

$$\forall x. (\forall x'. x' \prec_2 x \Rightarrow P(x')) \Rightarrow P(x) \qquad (3)$$

**Fact.** The induction step (3) implies $P(x)$ for all $x$ in normal form.

$$\forall x. \textit{ISN}(x) \Rightarrow P(x) \qquad (4)$$

**Proof.** By structural induction on $x$. The *At* case is easy. Since $\textit{If}(\textit{If}(u, v, w), y, z)$ is not normal, this case is vacuous. The $\textit{If}(At(a), y, z)$ case is

$$\textit{ISN}(\textit{If}(At(a), y, z)) \Rightarrow P(\textit{If}(At(a), y, z))$$

which simplifies, using the induction step, to

$$\textit{ISN}(y) \land \textit{ISN}(z) \Rightarrow P(y) \land P(z)$$

which follows from the induction hypotheses.

The Lemma is stated just like in section 4:

**Lemma.** $\forall xyz.\, P(y) \wedge P(z) \Rightarrow P(\mathit{If}(x, y, z))$

**Proof.** By structural induction on $x$. The $At$ case is proved as before. The $\mathit{If}$ case is

$$\forall yz.\, P(y) \wedge P(z) \Rightarrow P(\mathit{If}(\mathit{If}(u, v, w), y, z)) \,.$$

By (3) it is enough to show that $P(y)$ and $P(z)$ imply each of

$$P(\mathit{If}(v, y, z))$$
$$P(\mathit{If}(w, y, z))$$
$$\forall v'w'.\, \mathit{ISN}(v') \wedge \mathit{ISN}(w') \;\Rightarrow\; P(\mathit{If}(u, v', w'))$$

The induction hypotheses for $v$ and $w$ imply $P(\mathit{If}(v, y, z))$ and $P(\mathit{If}(w, y, z))$. It suffices to show $P(\mathit{If}(u, v', w'))$ for arbitrary normal expressions $v'$ and $w'$. The Fact (4) implies $P(v')$ and $P(w')$. Instantiate the induction hypothesis for $u$ with $y \to v'$ and $z \to w'$. Q.E.D.

The translation from the domain theory proof replaces $\mathit{ISN}(\mathit{norm}_2(x))$ by $P(x)$. The connection between the proofs is weaker that it was for $\mathit{norm}$. Domain theory allows explicit mention of $\mathit{norm}_2$'s recursive calls when instantiating $u$'s induction hypothesis; the recursion relation hides the calls via quantifiers.

The justification of $\mathit{norm}_2$ still requires simultaneous proofs that it terminates yielding a normal expression. The proof is by well-founded induction on $\prec_2$:

- Given $At(a)$ it makes no recursive calls and returns an atom, which is always normal.

- Given $\mathit{If}(At(a), y, z)$ it makes recursive calls on the predecessors $y$ and $z$. By induction hypotheses these calls return normal expressions so the final result is normal.

- Given $\mathit{If}(\mathit{If}(u, v, w), y, z)$ it makes recursive calls on predecessors $\mathit{If}(v, y, z)$ and $\mathit{If}(w, y, z)$. By induction hypotheses these return normal expressions $v'$ and $w'$. So $\mathit{If}(u, v', w')$ is a predecessor, justifying the final recursive call. By induction hypothesis this returns a normal expression.

This reasoning about the various cases of $\mathit{norm}_2$ can be formalized in my setting of well-founded recursion operators in Constructive Type Theory [10]. A function application has type $\sum_{y \in \mathit{exp}} \mathit{ISN}(y)$. It returns a *pair* of results: a normal expression $y$ and a proof object of type $\mathit{ISN}(y)$. Each recursive call on an argument $z$ must be justified by exhibiting a proof object of type $z \prec_2 x$. This is passed as an additional argument. In the *If-If* case, the outer call passes a proof object for $\mathit{If}(u, v', w') \prec_2$ $If(If(u, v, w), y, z)$, constructed from proof objects $ISN(u')$ and $ISN(v')$ from the inner calls.

After performing this elaborate construction of a well-founded recursion, the equations for $norm_2$ can be proved as usual in the approach [10].

# 9 Conclusions

Domain theory allows reasoning about recursion in a most flexible way, but at a heavy cost of complexity. The ideas can be difficult to grasp and introduce theoretical and practical obstacles. LCF is the only major theorem proving project that uses domain theory; even LCF users sometimes prefer to do without domains. The recursion relation proofs of *norm* and $norm_2$ suggest that domains are not essential for reasoning about many programs. Domain theory still has a vital role to play: there is no alternative for reasoning about compilers and continuously running processes.

Constructive Type Theory is concerned with terminating computations. Domain theory cannot be patched onto it; partial functions are completely antithetical to its view of computation. Using recursion relations it can express termination proofs of *norm* and $norm_2$, thereby deriving the function *normif*.

It is especially hard to prove the termination of functions involving *nested recursion*, since the termination of an outer recursive call may depend on a property of the results of the inner calls. The example $norm_2$ shows that this can be done without domains. Manna and Waldinger have studied a much more interesting nested recursive function, *UNIFY*, which performs unification [5]. They define a well-founded relation involving the structure of the expressions being unified and the number of distinct variables in those expressions. Under this relation, *UNIFY*'s outer recursive call can be justified only if its inner call returns a most-general, idempotent unifier of its arguments.

I formalized their work using the theorem prover Cambridge LCF, proving the total correctness of *UNIFY* [12]. The proof used a predicate *BEST_UNIFY_TRY* in a role analogous to that of *ISN* in the proof of $norm_2$: to allow the simultaneous proof of termination and correctness. It appears possible to verify *UNIFY* in Constructive Type Theory. Manna and Waldinger's well-founded relation is appropriate; there is no need to consider the recursion relation. *UNIFY* would return a substitution paired with a proof that this substitution had the necessary properties.

**Acknowledgements.** Robert Boyer and J Moore answered several queries about *norm* and generously hosted my visit to the University of Texas at Austin. Michael

J. C. Gordon read drafts of this paper.

# Appendix: the Cambridge LCF proof

Here is a sequence of commands that causes Cambridge LCF to prove the termination *norm* in domain theory. It is simpler than Leszczyłowski's Edinburgh LCF proof [4] because LCF has developed since then. No special ML code need be written: the data structure *exp* is defined automatically, and the standard rewriting tactic is powerful enough to handle both theorems. I performed this proof in half an hour at the terminal. I also proved the equivalence of *norm* and $norm_1$, and verified $norm_2$. These proofs are similar and not presented here.

Note that the ML used in Cambridge LCF is not Standard ML. An effort is underway to bring this ML up to date with the Standard.

This table gives LCF's printed representation of each logical connective:

| | | |
|---|---|---|
| `!` | $\forall$ | universal quantifier |
| `?` | $\exists$ | existential quantifier |
| `/\` | $\wedge$ | conjunction |
| `\/` | $\vee$ | disjunction |
| `==>` | $\Rightarrow$ | implication |
| `<=>` | $\iff$ | biconditional |
| `~` | $\neg$ | negation |
| `UU` | $\bot$ | undefined element of domain |
| `==` | $\equiv$ | equivalence (weak equality) |
| `<<` | $\sqsubseteq$ | partial ordering on domain |

The theory *exp* is declared, with a type operator of the same name. If $\tau$ is a type, then $\tau\, exp$ is the type of expressions whose atoms have type $\tau$. In LCF, *, **, ... are type variables. The `struct_axm` command defines expressions as a recursive type with strict constructors `ATOM` and `IF`. The constructor functions are *curried*: we must write `IF x y z` instead of $If(x, y, z)$.

```
new_theory 'exp';;
new_type 1 'exp';;
struct_axm (":* exp", 'strict',
    ['ATOM', ["a:*"]; 'IF', ["x: * exp"; "y: * exp"; "z: * exp"]]);;
```

The function symbol *norm* is declared, and a new axiom asserts its definition. The quantifier $\forall_D$ is not built in: we must write `!a.~ a==UU ==>` instead of $\forall_D a$.

```
new_constant ('NORM', ": (* exp) -> (* exp)");;

let NORM_CLAUSES =
new_closed_axiom ('NORM_CLAUSES',
 "(!a.~ a==UU : *  ==>  NORM(ATOM(a)) == ATOM(a))  /\
  (! a y z.~ a==UU:*  /\ ~ y==UU /\ ~ z==UU ==>
     NORM (IF (ATOM a) y z)  ==  IF (ATOM a) (NORM y) (NORM z))  /\
  (!u v w y z.~ u==UU /\ ~ v==UU /\ ~ w==UU /\ ~ y==UU /\ ~ z==UU ==>
     NORM (IF (IF u v w) y z)  ==
           NORM (IF u (IF v y z) (IF w y z)) : * exp)");;
```

An axiom, previously created by struct_axm, is bound to the ML identifier EXP_DEFINED. The structural induction tactic is instantiated to handle expressions and bound to the ML identifier EXP_TAC.

```
let EXP_DEFINED = axiom 'exp' 'DEFINED';;
let EXP_TAC = STRUCT_TAC 'exp' [];;
```

The Lemma is proved by induction followed by rewriting via the equations for *norm* and the totality of the constructors *At* and *If*. I have tweaked the statement of the Lemma to circumvent an annoyance involving admissibility of induction [3].

```
let NORM_LEMMA =
prove_thm ('NORM_LEMMA',
  "!x y z. ~ y==UU /\ ~ z==UU /\ ~ NORM(y)==UU /\ ~ NORM(z)==UU ==>
    ~ x==UU ==> ~ NORM(IF x y z)==UU : * exp",
  EXP_TAC "x" THEN
  ASM_REWRITE_TAC [NORM_CLAUSES; EXP_DEFINED]);;
```

The proof that *norm* is total resembles the proof of the Lemma.

```
let NORM_TOTAL =
prove_thm ('NORM_TOTAL',
  "!x. ~ x==UU ==> ~ NORM x ==UU : * exp",
  EXP_TAC "x" THEN
  ASM_REWRITE_TAC [NORM_CLAUSES; EXP_DEFINED; NORM_LEMMA]);;
```